\documentclass[12pt]{article}
\usepackage[cmtip,arrow]{xy}
\usepackage{pb-diagram,pb-xy}
\usepackage{amsmath}
\usepackage[psamsfonts]{amssymb}
\usepackage{cmmib57}
\usepackage{amsmath,amssymb,amscd}
\usepackage{tikz-cd}

\newcommand{\bPf}{\par\vspace*{-4pt}\indent{\sc Proof.}\enskip}
\newcommand{\ePf}{\medskip}
\def\QED{\hskip0.1em\hfill\null\ \null\nobreak\hfill\kern3pt\vbox{\hrule\hbox
   {\vrule\kern1pt\vbox{\kern1.7pt\hbox{$\scriptscriptstyle{QED}$}
    \kern0.2pt}\kern1pt\vrule}\hrule}}

\def\END{\hskip0.1em\hfill\null\ \null\nobreak\hfill\kern3pt\vbox{\hrule\hbox
   {\vrule\kern1pt\vbox{\kern1.7pt\hbox{$\,\,\,\vspace{5pt}$}
    \kern0.2pt}\kern1pt\vrule}\hrule}}
\newtheorem{theorem}{Theorem}[section]
\newtheorem{lemma}[theorem]{Lemma}
\newtheorem{corollary}[theorem]{Corollary}
\newtheorem{proposition}[theorem]{Proposition}
\newtheorem{remark}[theorem]{Remark}
\newtheorem{definition}[theorem]{Definition}
\newtheorem{example}[theorem]{Example}%---------------------------------------------------------------------%

\newcommand{\bCd}{\beq \begin{CD}}
\newcommand{\eCd}{\end{CD}\eEq}
\newcommand{\bcd}{\beq \begin{CD}}
\newcommand{\ecd}{\end{CD}\eeq}
\newcommand{\ben}{\begin{enumerate}}
\newcommand{\een}{\end{enumerate}}
\newcommand{\bEq}{\begin{eqnarray}}
\newcommand{\eEq}{\end{eqnarray}}
\newcommand{\beq}{\begin{eqnarray*}}
\newcommand{\eeq}{\end{eqnarray*}}
\newcommand{\bDf}{\begin{definition}\em}
\newcommand{\eDf}{\end{definition}}
\newcommand{\bLm}{\begin{lemma}}
\newcommand{\eLm}{\end{lemma}}
\newcommand{\bPr}{\begin{proposition}}
\newcommand{\ePr}{\end{proposition}}
\newcommand{\bTh}{\begin{theorem}}
\newcommand{\eTh}{\end{theorem}}
\newcommand{\bCr}{\begin{corollary}}
\newcommand{\eCr}{\end{corollary}}
\newcommand{\bRm}{\begin{remark}\em}
\newcommand{\eRm}{\end{remark}}
\newcommand{\bEx}{\begin{example}\em}
\newcommand{\eEx}{\end{example}}

\newcommand{\ie}{{\em i.e$.$} }
\newcommand{\eg}{{\em e.g$.$} }
\newcommand{\R}{I\!\!R}
\newcommand{\A}{\forall}

\newcommand{\der}{\partial}

\newcommand{\bxi}{\boldsymbol{\xi}}

\newcommand{\bp}{\boldsymbol{p}}

\newcommand{\bz}{\boldsymbol{z}}

\newcommand{\bE}{\boldsymbol{E}}

\newcommand{\bG}{\boldsymbol{G}}

\newcommand{\bM}{\boldsymbol{M}}
\newcommand{\bN}{\boldsymbol{N}}

\newcommand{\bP}{\boldsymbol{P}}

\newcommand{\bV}{\boldsymbol{V}}

\newcommand{\bZ}{\boldsymbol{Z}}
\newcommand{\wed}{\wedge}

\newcommand{\bet}{\beta}
\newcommand{\gam}{\gamma}

\newcommand{\tht}{\theta}

\newcommand{\lam}{\lambda}

\newcommand{\ome}{\omega}

\title{\large{Particle-like, dyx-coaxial and trix-coaxial Lie algebra structures  for a 
multi-dimensional continuous Toda type system
\footnote{Dedicated to Hartwig in occasion of his eighteenth birthday.}
}}

\author{ 
{\normalsize Marcella Palese\footnote{Corresponding Author} \, and Ekkehart Winterroth\footnote{Lepage Research Institute, 17. novembra 1, 
08116 Pre\v sov, Slovak Republic}}
\\ {\footnotesize Department of Mathematics,
University of Torino}
\\
{\footnotesize via C. Alberto 10, 10123 Torino, Italy} 
\\  {\footnotesize e--mail: 
{\sc marcella.palese@unito.it, ekkehart.winterroth@unito.it}}}
\date{}
\begin{document}
\maketitle

\begin{abstract}
We prove that with a $(2+1)$-dimensional Toda type system are associated algebraic skeletons which are (compatible assemblings)  of particle-like Lie algebras of dyons and triadons type.  We obtain trix-coaxial and dyx-coaxial Lie algebra structures for the system from algebraic skeletons of some particular choice for compatible associated absolute parallelisms. 
In particular, by a first choice of the absolute parallelism, we associate with the $(2+1)$-dimensional Toda type system a trix-coaxial Lie algebra structure 
made of two (compatible) base triadons constituting a $2$-catena.  Furthermore, by a second choice of the absolute parallelism, we associate a dyx-coaxial Lie algebra structure 
made of two (compatible) base dyons, as well as particle-like Lie algebra structures made of single $3$-dyons. Some explicit examples of applications such as conservation laws related to special solutions, and an inverse spectral problem are worked out.

\end{abstract}


%-------------------------------------------------------------
\section{Introduction}
%-------------------------------------------------------------

Toda type systems are nonlinear models which play a role in a variety of physical and, more in general, natural phenomena. 
 
The problem of integrability of nonlinear models has been recognized to be related to their algebraic properties in discrete and continuous, as well as, classical and quantum formulations.
 Algebraic properties can be interpreted as the counterpart of the concept of integrability given as of having `enough' conservation laws to exhaustively describe the underling field or associated dynamics.
Indeed, from an historical point of view, algebraic-geometric approaches are based on the requirement for the existence of conservation laws which emerge from internal symmetries (given in terms of algebraic structures).

In the Seventies, in fact, Wahlquist and Estabrook \cite{WaEs75,Estabrook} proposed a  technique for systematically deriving, from an integrable  system, what they called a `prolongation structure' in terms of a set of `pseudopotentials' related to the existence of an infinite set of associated conservation laws.
They also conjectured that, as a characterizing feature of the integrability property, the structure was `open' \ie not a set of structure relations of a finite--dimensional Lie group. Since then, `open' Lie algebras have been extensively studied in order to distinguish them from freely generated infinite-dimensional Lie algebras.  

Their interest in the study of integrability is in the fact that Lax pairs of the inverse spectral transform containing an isospectral parameter can be obtained by an homomorphism of the infinite-dimensional open Lie algebra in a finite-dimensional `closed' Lie algebra.
In their approach, conservation laws  are written in terms of `prolongation' forms and integrability is intended as a Frobenius
integrability condition for a `prolonged' ideal of differential forms describing intrinsically the given nonlinear model in the sense of \'E. Cartan.

Attempting a {\em description of symmetries in terms of Lie algebras} implies the appearance of an homogeneous space and thus the interpretation of prolongation forms as {\em Cartan--Ehresmann connections}.
It is clear that here the unknowns are both conservation laws and symmetries,
and the main point in this is how to realize the form of the conservation laws 
and thus the explicit expression of the prolongation forms. 
Different  prolongation ideals give rise to both different algebraic structures (symmetries) and corresponding conservation laws. 
By an inverse procedure based on the intrinsic duality between Lie algebras and differential systems \cite{Es82}, open Lie algebraic structures can `generate' whole families of different nonlinear systems bound by the same internal symmetry structure.

In a series of papers \cite{Pa05,Pa13,Pa16_Eco,PaWi02,PaWi10,PaWi11}, we explicated 
an algebraic-geometric interpretation of the above mentioned `prolongation' procedure 
in terms of {\em towers with infinitesimal algebraic skeletons} (in the sense of  \cite{Mo93}) and we will refer to that framework in this paper.  
It is noteworthy that slight modification of the internal symmetry properties generates {\em new} models which can contain possible integrable subcases.
For example, activator-substrate systems have been obtained by performing a slight modification of the internal symmetry algebra of twisted reaction-diffusion equations \cite{Pa16_Eco}.

The structure itself with which the tower forms are postulated can produce open algebraic structures or just Lie algebras. 
Our aim in this work is to investigate some common features of them and to show the emerging of particle-like Lie algebras structure as symmetry structures of integrable systems (associated with Poisson structures  the compatibility of which is worthy of study \cite{Fernandes93,Kos10}). 
Indeed, in general, infinite dimensional open Lie algebras are the main object of the search in view of the application of the inverse spectral transform to obtain soliton solutions, B\"acklund transformations and so on;  recent examples of applications can be found \eg in \cite{IgMa19,IgMa20,Morozov,Wang,Wang2,Li}. Although these features are not our prominent task in this paper, an inverse problem will be obtained in Section \ref{Lax section}.

Prolongation forms bringing to finite dimensional Lie algebras (without a spectral parameter) are generally discarded when searching for a Lax pair to be used within the inverse spectral transform. 

However, integrable systems, admitting infinite-dimensional prolongation Lie algebras can also admit finite-dimensional Lie algebras, which still can be related to some kind of internal symmetries of the systems themselves and to associated conservation laws, or even to B\"acklund transformations.
We refer, in particular, to the paper ``More prolongation structures'' by C. Hoenselaers \cite{Hoenselaers}, which pointed out two important features of the algebraic structures obtained by the method of Wahlquist and Estabrook. 

First feature: it can be that the prolongation forms can not always be solved in such a way that one obtains commutators among vector fields depending only on the `pseudopotentials' coordinates. A typical example is, in fact, the most general prolongation problem associated by such a procedure to equation \eqref{1}: in \cite{Palese} the prolongation problem was formally solved by introducing suitable operators of Bessel type, however a prolongation algebra (and then an inverse problem) could not be obtained explicitly. 

Furthermore, it is noteworthy that a certain arbitrariness is given by postulating the structure of the tower in the search of a skeleton. Say, to a given equation can be associated different towers with different skeletons. As it was already stressed by Estrabrook himself \cite{Es82} the same algebraic structure `contains' families of equations, associated linear spectral problems and B\"acklund transformations, interrelated by transformations between dependent and independent coordinates. 

The motivation of our previous research for the interpretation of prolongation structures as skeletons of towers was the aim of more deeply understanding such a feature: {\em equations, linear problems, B\"acklund transformations, are local coordinates expressions of common intrinsic structures}; this is of help also in practical questions: solutions of systems can be obtained by simpler systems having in common (part of) skeletons. In Section \ref{Conservation laws} we show that each one of the two compatible $4$-triadon constituting the skeleton given by a trix-coaxial Lie algebra structure generates the same conservation law and related special solutions.
This justifies the possible choice of a more restricted (instead of the most general one) form of the tower (then of the algebraic skeleton) still getting `solutions' (with this term meaning analytical solutions as well as particular conservation laws) of the original equation.

Second feature: even if the prolongation algebra is a finite dimensional (even abelian in his example) Lie algebra,  nevertheless there can exist B\"acklund transformations. It is shown that the exterior differential associated to the prolongation structure of the NLS equation being of genus $3$, and it is stressed that we can choose dependent and independent variables in an arbitray way. 
We can also lower the genus (in our skeleton formulation this means the choice of different representations $\rho$ or even different vector spaces $V$) so obtaining a reduced ideal where one of the independent coordinates is turned into a dependent coordinate. This turning a global symmetry in a local one provides Miura type transformations between the modified NLS and another system, the prolongation structure of which is finite dimensional and there is only one nontrivial potential entering a B\"acklund transformation acting on the modified NLS equation. 

In few words the Wahlquist-Estabrook method not always produces infinite dimensional open Lie algebras, but it could be that by that `procedure' we get  only a part of an algebraic skeleton. Therefore we can not automatically infer that, being the prolongation structure finite dimensional, then the system is not integrable. The results in \cite{Hoenselaers} are a counterexample, which suggest that we could extend a finite dimensional prolongation structure in order to implement the skeleton structure of an integrable family of nonlinear systems. These aspects are related to Olver's symmetry reduction \cite{Olver}.

Moreover, we guess that the skeleton can be further implemented by a finer structure related to particle-like Lie algebra structure and this is the inspiring idea of our investigations.
In this note we show that such (otherwise discarded) symmetries deserve a more careful study. We take a $(2+1)$-dimensional Toda type system as a study case and 
show that it posses algebraic properties related to the recently introduced concept of {\em particle-like Lie algebra structures} \cite{Vin17,Vin18}.

Vinogradov developed a completely abstract theory of compatibility of Lie algebra structures starting from the corresponding compatibility theory of Poisson structures.
Although the mathematical aspects of the theory are quite involved the nice point is that simple criteria of compatibility or non compatibility have been obtained which somehow have a certain grade of automatism. 

Furthermore, as for the physical side, Vinogradov speculated that this particle-like structures could be related to the ultimate particle structure of the matter: he noted that since
{\quote 
`the symmetry algebra $u(2) = so(3)$ of a nucleon can be assembled in one step
from three triadons [...] one might think that this structure of the symmetry reflects the fact that a nucleon is made from three ``quarks'' '. 
\quote}
This is of course only a speculation, but it also suggests a quite fascinating new perspective on internal symmetries of integrable systems.

%----------------------------------------------------------------------------------------------------------------------------------------------------
\section{Internal symmetries of Toda type systems in $(2+1)$ dimensions}
%----------------------------------------------------------------------------------------------------------------------------------------------------

Consider the $(2+1)$-dimensional system, a continuous (or long-wave) approximation of a spatially two-dimensional Toda lattice \cite{Toda}:
\bEq\label{1}
u_{xx}+u_{yy}+(e^{u})_{zz}=0\,,
\eEq
where $u=u(x,y,z)$ is a real field, $x,y,z$ are real local coordinates (if we want, $z$ playing the r\^ole of a `time') and the subscripts mean partial derivatives.
It can be seen as the limit for $\gam\to \infty$ of the more general model 
$$u_{xx}+u_{yy}+\left[\left(1+u/ \gam \right)^{\gam -1}\right]_{zz}=0$$ 
covering (for $\gam \neq 0,1$) various continuous approximations of  lattice models, among them the Fermi-Pasta-Ulam ($\gam =3$) \cite{Alfinito2}.
This model is almost {\em ubiquitus},  it appears in differential geometry; in mathematical and theoretical physics (Newman and Penrose);
in the theory of Hamiltonian systems; in general relativity; in the large $n$ limit of the $\textstyle{sl}(n)$ Toda lattice; in extended conformal 
symmetries, and theory of gravitational instantons;
in strings theory and statistical mechanics {\em etc.} (see \eg \cite{Boyer,Lebrun,Park,Plebanski}). 

It can be seen as the particular case with $d=1$ of so-called  $2d$-dimensional 
Toda-type systems \cite{Saveliev} obtained from a `continuum Lie algebra' by means of a zero curvature representation $u_{w\bar{w}}= K(e^{u})$, (in our particular case $w =x+iy$ and $K$ is  the differential operator  given by $K= \frac{\der^{2}}{\der z^{2}}$). 
In particular, it has been studied in the context of symmetry reductions \cite{Alfinito1,Grassi} and a $(1+1)$-dimensional version in the context of prolongation structures \cite{Alfinito2}. The $(2+1)$-dimensional system has been associated with a Ka\v c--Moody Lie algebra and related to Saveliev's continuum Lie algebras of particular kind \cite{PaWi10}.

The Toda system  \eqref{1} can be put in the complex form
\beq
\der_\zeta\der_{\bar\zeta} u =  -1/4 \der^{2}_{z} e^{u} \,,
\eeq
by the transformations
$\zeta = g(\eta)$, $\bar{\zeta} = \bar{g}(\bar{\eta})$, $u = \tilde{u}
- \ln(g^{'}\bar{g}^{'})$,
where $\zeta = x+ iy$,  $\bar\zeta = x- iy$, $\der_\zeta =\frac{1}{2}
(\der_x - i \der_y)$, $\der_{\bar\zeta}=\frac{1}{2}
(\der_x + i \der_y)$, $g^{\prime}=g_{\eta}(\eta)$, 
$\bar{g}^{'}=g_{\bar\eta}(\bar\eta)$ and $g(\eta)$ is an arbitrary 
holomorphic function of $\eta= x^{\prime}+iy^{\prime}$.
A Lax pair for this complex form of the $2D$ Toda equation has been found; see \eg Manakov and Santini \cite{MaSa09} and references therein; original references are  \cite{Saveliev}, as well as \cite{Zakharov}.

%----------------------------------------------------------------------
\subsection{Skeletons for the (2+1) Toda system}
%----------------------------------------------------------------------

Let us first recall a few mathematical tools constituting the background for a detailed treatment of which 
we refer to \cite{PaWi02,PaWi10,PaWi11} and \cite{Mo93,PRS79}.  

From one side global properties of partial differential equations such as internal symmetries and invariance properties having an issue in dynamics can be described by mathematical tools which enable us to deal with global properties at large scales, connecting local data to global ones.
On the other side transformations of configurations of a system can be globally studied by means of the theory of the action of Lie groups on manifolds.
The differential content carried by a Lie group (and its Lie algebra) and by its structure equations provides differential equations. 

We observe that two ingredients constitute the  nonlinear phenomena: symmetries on the one side (algebraic content) and changes in time and space on the other side (differential content).
In particular, to keep account of the `interaction' of both aspects, we recognize a refined structure of open Lie algebraic structures associated with them: we introduce
a notion which generalizes the concept of a homogeneous space, \ie that of an {\em algebraic skeleton} $\bE = \mathfrak{g} \oplus \bV$ on a finite-dimensional vector space $\bV$, with $\mathfrak{g} $ a possibly infinite dimensional Lie algebra. The further step is introducing a tower with such a skeleton.

An {\em algebraic skeleton} on a finite-dimensional vector space $\bV$ is a triple $(\bE,\bG,\rho)$, with $\bG$ a (possibly infinite-dimensional) Lie group, 
$\bE = \mathfrak{g} \oplus \bV$ is a (possibly infinite-dimensional) vector space {\em not necessarily equipped with a Lie algebra structure}, 
$\mathfrak{g}$ is the Lie algebra of $\bG$, 
and 
$\rho$ is a representation of $\mathfrak{g}$ on $\bE$ such that it reduces to the adjoint representation of $\mathfrak{g}$ on itself. The fact that $\bE$ is not a direct sum of Lie algebras, but an open algebraic structure is fundamental in order to be able to generate whole families of nonlinear differential systems, starting from it.

We now consider a suitably constructed differentiable structure which is somewhat modelled on the skeleton above. Let us introduce a differentiable manifold $\bP$ on which a Lie group $\bG$, with Lie algebra $\mathfrak{g}$, acts on the right; $\bP$ is a principal bundle $\bP\to\bZ\simeq\bP/\bG$. 
By construction, we have that $\bZ$ is a manifold of type
$\bV$, \ie $\A \bz\in \bZ$,  $T_{\bz} \bZ \simeq \bV$. 

Suppose we have a way to define a representation $\rho$ of the Lie algebra $\mathfrak{g}$ on $T_{\bz} \bZ \simeq \bV$, in such a way that it could be possible under certain conditions to find a %
homomorphism between the open infinite dimensional Lie algebra, constructed by $\rho$, and a quotient Lie algebra.
Let us call $\mathfrak{k}$ the (possibly infinite dimensional) Lie algebra obtained as the direct sum of  such a quotient Lie algebra  with $\mathfrak{g}$. 
From the differentiable side, a {\em tower $\bP(\bZ,\bG)$ on $\bZ$ with skeleton $(\bE, \bG,\rho)$ } is 
an {\em  absolute parallelism} $\ome$ on $\bP$ valued in $\bE$, 
invariant with respect to $\rho$ and reproducing elements of $\mathfrak{g}$ from the fundamental
vector fields induced on $\bP$, \ie  $R^{*}_{g}\omega = \rho(g)^{-1}\ome$, for
$g\in \bG$; $\ome(\tilde{A}) = A$, for $A\in\mathfrak{g}$;
here $R_{g}$ denotes the right translation and $\tilde{A}$ the fundamental
vector field induced on $\bP$ from $A$.
In general, the absolute parallelism  {\em does not}  define a Lie algebra homomorphism.
 
Let then $\mathfrak{k}$ be a Lie algebra and $\mathfrak{g}$ a Lie subalgebra of
$\mathfrak{k}$. 
Let $\bG$ be a Lie group with Lie algebra $\mathfrak{g}$ and $\bP(\bZ, \bG)$ be a principal fiber bundle with structure group $\bG$ over a manifold $\bZ$ as above. 
A {\em Cartan connection} in $\bP$ of type
$(\mathfrak{k}, \bG)$ is 
a $1$--form $\ome$ on $\bP$ with values in
$\mathfrak{k}$ such that
$\ome |_{T_{\bp} \bP}: T_{\bp} \bP\to \mathfrak{k}$ is an isomorphism $\forall \bp \in
\bP$, 
$R^{*}_{g}\ome=Ad(g)^{-1}\ome$ for $g\in \bG$ 
and 
reproducing elements of $\mathfrak{g}$ from the fundamental
vector fields induced on $\bP$.
It is clear that a Cartan connection
$(\bP, \bZ, \bG, \ome)$ of type $(\mathfrak{k}, \bG)$ is a special case of a tower on $\bZ$.

The vector space $\bV$ is finite dimensional and generated by some of the vector fields in the prolongation structure. It has the property that each  bracket of some of remaining vector fields of the prolongation structure (freely generating an infinite dimensional Lie algebra $\mathfrak{g}$) with its generators is again in $\bV$. 
In particular unknown commutators in the freely generated Lie algebra are related in such a way that their assigned relations are elements of $\bV$.

As an example of application of such an abstract formulation to the real world we refer  \eg to \cite{Pa16_Eco}, whereby activator-substrate systems have been obtained by performing a slight modification of the internal symmetry algebra of twisted reaction-diffusion equations: the necessary condition for the generation of stable patterns (related to general integrability properties in the limit of a null normalized diffusion constant) are  formulated in terms of `closeness' properties within the symmetry algebra vector space.

Following \cite{PaWi10}, we recall how to get both some skeletons and towers over them associated with the system \eqref{1}. 

On a manifold with local coordinates
$(x,y,z,u,p,q,r)$, we introduce the closed differential ideal defined by the set of $3$--forms:
$
\theta_{1} = du \wedge dx \wedge dy - rdx \wedge dy \wedge 
dz$, 
$\theta_{2} = du \wedge dy \wedge dz - pdx \wedge dy \wedge 
dz$,
$\theta_{3} = du \wedge dx \wedge dz + qdx \wedge dy \wedge 
dz$,
$\theta_{4} = dp \wedge dy \wedge dz - dq \wedge dx \wedge 
dz$ $ +$ $ e^{u}dr\wedge dx \wedge dy + e^{u}r^{2}dx \wedge dy
\wedge dz$. 
It is easy to verify that on every integral submanifold defined by $u = u(x,y,z)$, 
$p=u_{x}$, $q=u_{y}$, $r=u_{z}$, with $dx\wedge dy\wedge dz\neq 0$, the above ideal
is equivalent to the Toda system under study. 

By an ansatz first introduced in \cite{PaWi02}, we  look for suitable $2$--forms (generating associated conservation laws)
\beq
\Omega^{k}= \tht^k_m\wed\ome^m 
\eeq 
where 
$\tht^k_m = - \hat{A}^{k}_{m} dx-\hat{B}^{k}_{m}dy- \hat{C}^{k}_{m}dz$, with  $\hat{A}^{k}_{m}$, $\hat{B}^{k}_{m}$, $\hat{C}^{k}_{m}$ elements of $N \times N$ 
constant regular matrices, and the absolute parallelism forms are given by 
\bEq\label{omega}
\ome^m= d\hat{\xi}^m+ \hat{F}^m dx  +\hat{G}^m dy  +  \hat{H}^m dz\,,
\eEq
\ie
\bEq \label{Omega}
\Omega^{k} = H^{k}(u, u_{x}, u_{y}, u_{z}; \xi^{m})dx \wedge dy+
F^{k}(u, u_{x},u_{y}, u_{z}; \xi^{m})dx \wedge dz + \\
+ G^{k}(u, u_{x}, u_{y}, u_{z}; \xi^{m})dy \wedge dz+ A^{k}_{m}d \xi^{m} \wedge dx
+ B^{k}_{m}d \xi^{m} \wedge dz + d \xi^{k}
\wedge dy\,, \nonumber
\eEq 
where $\xi=\{\xi^{m}\}$, $k,m=1,2,\ldots,
{\rm N}$ (N arbitrary), and $H^{k}$, $F^{k}$ and $G^{k}$ are, respectively, the
pseudopotentials and functions to be determined, while $A^{k}_{m}$ and $B^{k}_{m}$ denote the elements of two $N \times N$ 
constant regular matrices related to the previous ones and we have rescaled the coordinates $\xi^k$.
In particular note that (see also \cite{PaWi02,Pa05})
\bEq 
F^k = \hat{C}^{k}_{m}\hat{F}^m- \hat{A}^{k}_{m}  \hat{H}^m  \label{relations1} \,, \\
G^k = \hat{C}^{k}_{m}\hat{G}^m- \hat{B}^{k}_{m}  \hat{H}^m  \label{relations2} \,, \\
H^k = \hat{B}^{k}_{m}\hat{F}^m- \hat{A}^{k}_{m}  \hat{G}^m \label{relations3} \,, \\
\xi^k = \hat{C}^{k}_{m}\hat{\xi}^m \,.
\eEq

The integrability  condition for the ideal generated by forms 
$\theta_{j}$ and $\Omega^{k}$ finally yields
\bEq\label{H}
H^{k} = e^{u}u_{z}L^{k}(\xi^{m}) + P^{k}(u,\xi^{m}) \,,
\eEq
\bEq\label{F}
F^{k} = -\,u_{y}L^{k}(\xi^{m}) + Q^{k}(u, \xi^{m}) \,,
\eEq
\bEq\label{G}
G^{k} = u_{x}L^{k}(\xi^{m}) + M^{k}(u,\xi^{m}) \,,
\eEq
where $L^{k}$, $P^{k}$, $Q^{k}$, $M^{k}$ are functions of integration. 

It turns out that $Q^{k}(u, \xi^{m})$ can be written in terms of the others. 
Indeed we have (see \eg \cite{Morris,Tondo})
$H^{k}  =A^{k}_{m} G^{m} - B^{k}_{m} F^{m} $
so that $ e^{u}u_{z}L^{k}(\xi^{l}) + P^{k}(u,\xi^{l})=A^{k}_{m} (u_{x}L^{m}(\xi^{l}) + M^{m}(u,\xi^{l})) - B^{k}_{m}(-\,u_{y}L^{m}(\xi^{l}) + Q^{m}(u, \xi^{l}))$, \ie
$B^{k}_{m}Q_u^{m}(u, \xi^{l})= e^{u}u_{z}L^{k}(\xi^{l}) + P_u^{k}(u,\xi^{l})+ A^{k}_{m}  M_u^{m}(u,\xi^{l})$, which can be integrated once the dependence on of $P^{k}(u,\xi^{l})$ and $M^{k}(u,\xi^{l})$ on $u$ is given.
As a consequence, the desired representation $\rho$ for the skeleton is provided by the following equations (we omit the indices for simplicity) \cite{Palese,PaWi10}.
\bEq
P_{u} = e^{u}[L,M]\,,\label{P} \, \quad
 M_{u} = - [L,P]\,,\label{M} \, \quad 
 \left[M,P\right] = 0\,.\label{comm}
\eEq
Note that here  $L$ depends only on $\xi^{m}$, while $P$ and $M$ still have a dependence on $u$ determined by the first two differential equations. A tower with $P$ and $M$ given in terms of $L$ has been obtained by suitable {\em operator Bessel coefficients} \cite{Palese}.

Note that formally this tower shall provide the Lax pair of an inverse spectral problem; however, it is a non trivial task to characterize explicitly its algebraic skeleton by means of the representation provided by  the relations $\left[M,P\right] = 0$, \ie to obtain a spectral problem in a manageable form.

Particular choices for the absolute parallelism can provide us explicit representations of the prolongation skeleton; in particular a Ka\v c--Moody Lie algebra has been obtained \cite{PaWi10} (see Proposition \ref{dyonsProposition}, case $1. \, (b)$ below). 
In the following we will concentrate on those choices that generate particle-like Lie algebra structures.
We shall see that it is yet possible to obtain a spectral problem with a particular choice of the tower.

%-------------------------------------------------------------------------------------------
\section{Particle-like Lie algebra structure}
%-------------------------------------------------------------------------------------------

Recently, Vinogradov proved that any Lie algebra over an algebraically closed field or over $\R$ can
be assembled in a number of steps from two elementary constituents, that he called dyons
and triadons \cite{Vin17}. He  considered the problems of the construction and classification of
those Lie algebras which can be assembled in one step from base dyons and triadons,
called coaxial Lie algebras. The base dyons and triadons are Lie algebra structures
that have only one non-trivial structure constant in a given basis, while coaxial Lie
algebras are linear combinations of pairwise {\em compatible} base dyons and triadons \cite{Vin18}. Here for the convenience of the reader we recall some basic facts of the theory in the original Vnogradov's notation.

\bDf
Lie algebra structures $\mathfrak{g}_1$  and $\mathfrak{g}_2$ on a vector space $ V$ are called compatible
if $[,]_{\mathfrak{g}_1}+[,]_{\mathfrak{g}_2}$ is also a Lie algebra product.

A Lie algebra $\mathfrak{g}$ is called simply assembled from Lie algebra structures $\mathfrak{g}_1,\dots,\mathfrak{g}_m$ on $|\mathfrak{g}|=V$ if
the Lie algebras  $\mathfrak{g}_i$'s are pairwise compatible and
$[,]_{\mathfrak{g}}=[,]_{\mathfrak{g}_1}+\dots [,]_{\mathfrak{g}_m}$ .
Note that if the Lie algebras $\mathfrak{g}_i$'s are compatible, then any linear combination of compatible Lie algebras commutators is a Lie  algebra commutator (or product) .
\eDf

\bDf
Fix a basis $B=e_1,\dots e _n$ in the representation vector space of a given Lie algebra. Let $i,j,$ and $k$ be integers, $1\leq i,j,k \leq n$, no two of them equal,
and denote by $\{i,j | k\}$ (respectively, $\{i | j\}$) the $n$-triadon (respectively, the $n$-dyon) such that $[e_i,e_j] = - [e_j,e_i] = e_k$ (respectively, $[e_i,e_j] = - [e_j,e_i] = e_j$) are the only non-trivial Lie commutators of basis vectors. 
Vinogradov called  them  `base triadon' and `base dyon', respectively
or by the unifying term `base lieon'.

An $n$-dyon is the direct sum of a dyon with an $n-2$-dimensional abelian Lie algebra, $n\geq 2$, (\ie there is only one  non vanishing bracket and it is a dyon).
Analogously an $n$-triadon is the direct sum of a triadon with an $n-3$-dimensional abelian Lie algebra, $n\geq 3$ (\ie there is only one  non vanishing bracket and it is a triadon). They can also be referred generically as $n$-lieons. 

A linear combination of pairwise compatible base lieons is called a {\em coaxial Lie algebra structure}. 
A Lie algebra structure will be called {\em trix-coaxial} (respectively, {\em dyx-coaxial}) if it consists only of base triadons (respectively,
base dyons).
A coaxial Lie algebra $\mathfrak{g}$ may be presented as a linear combination,
\beq 
 \mathfrak{g}=\sum{}{} \alpha_{(i,j|k)} \{i, j|k\} + \sum{}{} \bet_{(m|n)} \{m|n\} 
\eeq
of pairwise compatible base lieons.

The vectors $e_i,e_j $, and $e_k$ (respectively, $e_i,e_j$) are called the {\em vertices of the triadon} $\{i,j | k\}$
(respectively, {\em of the dyon }$\{i | j\}$). The vectors $e_i$ and $e_j$ are called  the {\em ends of the triadon} $\{i,j | k\}$, while  $e_k$ is the {\em center of the triadon}.
The {\em origin and the end  of  the dyon $\{i | j\}$} are $e_i$ and $e_j$, respectively.  The base
triadons $\{i,j | k\}$ and $\{j,i | k\}$= - $\{i,j | k\}$ are not distinguished since they have identical compatibility properties.
\eDf
We now recall Proposition $3.1$ of \cite{Vin18} stating some necessary and sufficient conditions for the compatibility or incompatibility  of particle-like Lie algebra structures:
\begin{itemize}
\item Two base triadons are non-trivially compatible if and only if they have a
common center, a common end, or both.
\item Two base dyons are incompatible if and only if the origin of one is the end of the other and
they have no other common vertices.
\item A base dyon is non-trivially compatible with a base triadon if and only if its origin coincides
with one of the ends of the triadon.
\end{itemize}
For further notation and vocabulary we refer the reader to Vinogrados's papers.

%--------------------------------------------------------------------------------------------------------------------------------------------------------
\subsection{Trix-coaxial, dyx-coaxial and particle-like Lie algebra structures for the Toda system}
%--------------------------------------------------------------------------------------------------------------------------------------------------------

We prove that with a $(2+1)$-dimensional Toda type system are associated algebraic skeletons which are compatible  assemblings  of particle-like Lie algebras of dyons and triadons type.  We obtain trix-coaxial and dyx-coaxial Lie algebra structures for the system from skeletons of some particular choice for compatible associated absolute parallelisms. In particular, we find a trix-coaxial Lie algebra structure 
made of two (compatible) base triadons constituting a $2$-catena (see Proposition 3.1, pag 5 \cite{Vin18}). 

Let us indeed now look for special skeletons.

%-------------------------------------------------------------------
\subsection{Trix-coaxial  Lie algebra structures}
%-------------------------------------------------------------------

\bPr
Associate with the Toda type system  \eqref{1} is a trix-coaxial Lie algebra structure made of two (compatible) base triadons constituting a $2$-catena.
\ePr

\bPf
If we look for operators  $P(u,\xi)= e^{u}\bar{P}(\xi)$, $M(u,\xi)=M(e^u,\xi)$,  we get
$M( e^u;\xi) $ $=$ $ - e^u[L(\xi),\bar{P}(\xi)]+ \bar{M}(\xi)$ and thus
$\bar{P}(\xi)$ $=$ $ - e^u[L(\xi),[L(\xi),\bar{P}(\xi)]]+[L(\xi),\bar{M}(\xi)]$.
There are additional relations determined by the third prolongation equation 
$[- e^u[L(\xi),\bar{P}(\xi)]+ \bar{M}(\xi), e^u \bar{P}(\xi)]=0$.

Let us then put $L=X_1,\bar{M}=X_2, \bar{P}=X_3, [X_1 ,X_3 ]= X_4 $. From the above we  have the following prolongation closed Lie algebra
\beq
[X_1 ,X_2 ]=X_3 \,,   [X_1 ,X_3 ]= X_4 \, , [X_1 , X_4 ]= [X_2 ,X_3 ]= [X_2 ,X_4 ]= [X_3 ,X_4 ]= 0 \,.
\eeq
The above is a trix-coaxial Lie algebra structure made of two compatible $4$-triadons. 

Indeed, 
by taking $X_4=0$, we get 
 $ [X_1 ,X_2 ]=X_3 \,,   [X_1 ,X_3 ]=   [X_2 ,X_3 ]= 0$ and  $ [X_1 , X_4 ]=   [X_2 ,X_4 ]= [X_3 ,X_4 ]= 0$ trivially.
 
On the other hand by taking $X_2=0$, we get 
$[X_1 ,X_3 ]= X_4 \,, [X_1 , X_4 ]=   [X_3 ,X_4 ]= 0$ and  $[X_1 ,X_2 ]= [X_2 ,X_3 ]= [X_2 ,X_4 ] =0 $ trivially.

According with \cite{Vin18} the two $4$-triadons above {\em are non trivially compatible having a common end $X_1$}, and they constitute a $2$-catena.
\ePf

%-------------------------------------------------------------------------------------------------------------------------------------------
\subsubsection{Conservation laws and special solutions associated with a $2$-catena}\label{Conservation laws}
%-------------------------------------------------------------------------------------------------------------------------------------------

Let us now explicate the tower corresponding to such $4$-triadons.

For the sake of simplicity let us put $A^{k}_{m}=B^{k}_{m}=\delta^{k}_{m}$, were $\delta^{k}_{m}$ is the Kronecker symbol. 
By substituting the above commutators into  equations \eqref{H} and \eqref{G} (the expression of \eqref{F} being constrained in this case by the relation $F=G+H$), we get

\bEq\label{H2}
H = e^{u}u_{z}X_1+ e^{u} X_3 \,,
\eEq
\bEq\label{G2}
G = u_{x}X_1 - e^{u} [ X_1, X_3]+ X_2 \,,
\eEq

Now from equation \eqref{Omega}, by sectioning we obtain

\bEq\label{boh}
H^{k}  -  \xi^k_y = -  \xi^k_x \,,
\eEq
\bEq\label{boh2}
G^{k} + \xi^k_y =  \xi^k_z \,,
\eEq
(together with $F^{k} + \xi^k_x =  \xi^k_z $ which depends on the two others).

We note that each one of  $4$-triadons above can be represented in a space of local coordinates $\xi^k$ providing conservation laws related to two compatible Poisson structures. 

Indeed let us consider the $4$-triadon given by $X_2=0$. A representation in the coordinates $\{\xi^1,\xi^2,\xi^3\}$ is given by 
$X_1= \xi^2 \der/\der \xi^1$, $X_3= \xi^3 \der/\der \xi^2$ and $X_4= -\xi^3 \der/\der \xi^1$.

The tower corresponding to this case gives 
\beq\label{H2'}
 e^{u}u_{z}\xi^2 \der/\der \xi^1+ e^{u}  \xi^3 \der/\der \xi^2  -  \xi^k_y \der/\der \xi^k= -  \xi^k_x \der/\der \xi^k  \,,
\eeq
\beq\label{G2'}
 u_{x}\xi^2 \der/\der \xi^1 + e^{u} \xi^3 \der/\der \xi^1+ \xi^k_y\der/\der \xi^k =  \xi^k_z\der/\der \xi^k\,,
\eeq
which gives us the system
\bEq
\bxi_x= \bxi_y +\bM \bxi \, \\
\bxi_z= \bxi_y + \bN \bxi \,
\eEq
where $\bxi =(\xi^1,\xi^2,\xi^3)^T$,  $\bM$ and $\bN$ are $3\times 3$ matrices such that $M_{12}= -e^u u_z$,$ M_{23}= -e^u$, $N_{12}=u_x$, $N_{13}=e^u$, and all the other entries are zeros. 
In view of \eqref{relations1}-\eqref{relations3} and \eqref{omega} this system can be interpreted as a conservation law.
It is a  first order system which can be manipulated by resorting to the method of characteristics so that it turns out to be useful to find out special solutions of the Toda system.

Note indeed that this system is equivalent to the following system of coupled equations of Maxwell type 
\bEq
& & \xi^1_{yy} - \xi^1_{xx} = (e^u u_z\xi^2)_x +(e^u u_z\xi^2)_y \,, \\
& & \xi^2_{yy} - \xi^2_{xx} = (e^u \xi^3)_x +(e^u \xi^3)_y \,,  \label{second eq} \\
& & \xi^1_{zz} - \xi^1_{yy} =  (e^u \xi^3 +u_x\xi^2)_z + (e^u \xi^3 +u_x\xi^2)_y\,,
\eEq
where $e^u u_z\xi^2$, $e^u \xi^3$ and $e^u \xi^3 +u_x\xi^2$ can be recognized as charge/current densities.
On the other hand, the system can be simplified since $\xi^2_y = \xi^2_z$; we have that \eqref{second eq} can also be written as 
\bEq
\xi^2_{zz} - \xi^2_{xx} = (e^u \xi^3)_x +(e^u \xi^3)_y \,, 
\eEq
from which we obtain the Maxwell-type equation 
\bEq
\xi^2_{yy} = \xi^2_{zz}  \,.
\eEq
Further  manipulations can be made by using $\xi^3_y = \xi^3_y = \xi^3_z$.

We remark that the same conservation law and related outcomes are obtained by the $4$-triadon given by $X_4=0$.
Therefore existence of that tower with a finite dimensional skeleton which is a $2$-catena says us that the two Poisson structures corresponding to each $4$-triadon are compatible also in a sense which is interpretable from a physical point of view: they are structures associated with the same Toda system,  and more precisely with the same conservation law and related special solutions.
Compatibility of Poisson structures is beyond the scope of this paper, however this result suggest interesting links between special solutions and compatible Poisson structures, which will be the object of further investigations (in particular for meron-like configurations or gravitational instantons).

%------------------------------------------------------------------------------------------
\subsection{Dyx-coaxial and particle-like Lie algebra structures}
%------------------------------------------------------------------------------------------

In the following we analyze with more detail the case of choice  $P(u,\xi)= \ln u \bar{P}(\xi)$, $M(u,\xi)=M(e^u,\xi)$ studied in \cite{PaWi10} also leading to an infinite dimensional skeleton homomorphic to a Ka\v{c}-Moody Lie algebra. We carefully distinguish the various cases.

This choice of the absolute parallelism associates with the Toda system \eqref{1} dyx-coaxial  and particle-like Lie algebra structures.

First we need a preliminary result (see also \cite{PaWi10}).
\bLm\label{lemma}
Let $P(u,\xi)= \ln u \bar{P}(\xi)$, $M(u,\xi)=M(e^u,\xi)$. We get the following infinitesimal algebraic  skeleton with the structure of an {\em open Lie algebra}:
\bEq\label{open}
[X_1, X_2]=X_4\,, [X_1, X_3]=X_5\,, [X_4, X_5]  
=[X_2,X_7] \,, [X_3, X_4] =  [X_2, X_5] \,,
\eEq
\beq
[X_1, X_4]=X_6\,,  [X_1, X_5]=X_7\,, [X_2, X_3] = X_8  \,, 
\eeq
\beq
[X_1,X_8]= [X_2, X_4] =  [X_2, X_6] =[X_3, X_7] = 0\,,
\ldots
\eeq
\eLm

\bPf
Put  $L=X_1 (\xi)$.

By derivation
we get
$M(e^u,\xi)=- (\ln u -1)u[X_1 (\xi),X_3(\xi)]+X_2 (\xi)$,
and 
$P(u,\xi)= ue^u \ln u[X_1(\xi) , M] $.

For $u \neq 0,1$ (which are trivial solutions of the Toda system), from $[P,M]=0$
we get
\beq
[[X_1, M],M]=0\,;
\eeq
from which we get 
\beq
[[X_1,X_2],X_2 ] = 0 \,, 
\,
[X_1,[X_1 ,X_3 ]], X_2 ]+ [[X_1 ,X_2 ],[X_1 ,X_3 ]] =0\,,
\eeq
\beq
[[X_1,[X_1 ,X_3 ]],[X_1 ,X_3]] = 0\,.
\eeq
By putting 
$[X_1, X_2]=X_4$, $[X_1, X_3]=X_5$, $[X_1, X_4]=X_6$, $[X_1, X_5]=X_7$, $[X_2, X_3] = X_8$, we obtain an infinite dimensional skeleton as follows
\beq
 [X_1, X_8]= [X_2, X_4] =  [X_2, X_6] =[X_3, X_7] = 0\,,
 \eeq
 \bEq\label{commutator relations}
  [X_4, X_5] =[X_2,X_7] \,, [X_3, X_4] =  [X_2, X_5] \,, 
   \eEq
 \beq
\ldots
\eeq
\ePf
Here the dots means that we can continue this structure by introducing new generators still obtaining the peculiar relations  of the type \eqref{commutator relations} which distinguish this algebraic structure from a freely generated Lie algebra (see the discussion in \cite{Pa16_Eco}).

\bPr\label{dyonsProposition}
The homomorphism  $X_4=\lam X_2$ and $X_5 =\mu X_3$ associates with the Toda system \eqref{1} dyx-coaxial and particle-like Lie algebra structures   as well as an infinite-dimensional Lie algebra homomorphic with a Ka\v{c}-Moody Lie algebra.
\ePr
 
\bPf
We essentially distinguish the two cases $X_8\neq 0$ and $X_8 = 0$, together with various different subcases.
\begin{enumerate}
\item
if $[X_2, X_3] = X_8\neq 0$, then $\mu = -\lam$ must old; we can distinguish different subcases
 \begin{enumerate}
 \item in general  the case $X_8\neq 0$ and $\mu= -\lam \neq 0$ can provide infinite-dimensional Lie algebras homomorphic with Ka\v{c}-Moody type Lie algebras.
\item  the particular case $X_8=\nu X_3$ and $\mu = -\lam = 1$ (\ie $X_4=X_2$ and $X_5 = -X_3$) giving an infinite-dimensional Lie algebra homomorphic with a Ka\v{c}-Moody Lie algebra was obtained in \cite{PaWi10}.
 \item 
 the particular case $X_8=\nu X_3$ and  $\mu= -\lam = 0$ (\ie $X_4=X_5=0$; see \cite{PaWi10})  gives a particle-like Lie algebra as a base $3$-dyon:
\bEq\label{dyon}
[X_1, X_2]=0\,, [X_1, X_3]=0\,, [X_2, X_3] = \nu X_3\,.
\eEq
\end{enumerate}
\item  if  $[X_2, X_3] = X_8 = 0$, 
then $X_6=X_4$, $X_7=X_5$, and we distinguish the following different subcases (the case $\mu= \lam = 0$ giving an abelian Lie algebra):
 \begin{enumerate}
 \item
the case $\mu=0$ and $\lam\neq 0$ provides us with a particle-like Lie algebra as a base $3$-dyon:
\bEq\label{3-dyon}
[X_1, X_2]=\lam X_2\,, [X_1, X_3]=0\,, [X_2, X_3]  =0 \,.
\eEq
\item
the case
$\lam =\mu \neq 0$ provides a  dyx-coaxial Lie algebra structure as an assembling of two compatible base  $3$-dyons
\bEq\label{dyx-family}
[X_1, X_2] = \lam X_2\,, [X_1, X_3] =  \lam X_3\,, [X_2, X_3] = 0 \,.
\eEq
 \item the particular case  $\lam =\mu=1$ (\ie $X_4=X_2$ and $X_5 =X_3$) gives
\beq
[X_1, X_2]=X_2\,, [X_1, X_3]=X_3\,, 
[X_2, X_3] = 0 \,,
\eeq
and it  was obtained in  \cite{PaWi10}.
\end{enumerate}
\end{enumerate}
\ePf

\bRm
We can try to check the compatibility of the above particle-like  Lie algebra structures.
The latter Lie algebra \eqref{dyx-family} is constituted of two mutually compatible dyons (dyx-coaxial Lie algebra), whose the first is given by \eqref{3-dyon}, while the Lie algebras \eqref{dyon} and \eqref{3-dyon} are made of a single dyon and they are not compatible.
Indeed we note that the first dyon of \eqref{dyx-family} is not compatible with \eqref{dyon}, while the second dyon of \eqref{dyx-family} is. 

The question now is can we still construct a different  dyx-coaxial Lie algebra  from the original skeleton? For example the following would be a dyx-coaxial Lie algebra of compatible dyons
\bEq
[X_1, X_3] =  \lam X_3, [X_2, X_3] = \nu X_3, [X_1, X_2] = 0 \,
\eEq
We ask if we can get it from the prolongation skeleton by a suitable quotienting, \ie if it is somehow compatible with (or derivable from) the skeleton structure. 
However, we note that if we put $X_4=0$ from the beginning (which is the case when we assume that  $ [X_1, X_2] = 0$), and  if  $X_8 =\nu X_3$, this would  imply also $[X_1, X_3] = 0$, then we  would get \eqref{dyon} back. 

Thus it appears that the case $X_8 =\nu X_3$ corresponds or to a particle-like Lie algebra structure or to a Ka\v{c}-Moody type Lie algebra (it is noteworthy that the latter is anyway  an infinite-dimensional loop Lie algebra of a dyx-coaxial Lie algebra) and these two cases appear to be non compatible.

Let us then investigate from a more general point of view this feature.
We ask whether  we can look for different quotient homomorphisms.

Let now consider the case $X_4 = 0$ from the beginning, and $X_8\neq 0$, and look for a quotient lie algebra given by
$X_8 = - \gamma X_2$, , $X_5=  \mu X_3$, and we obtain the  Lie algebra structure depending on two parameters 
\bEq\label{incompatible}
[X_1, X_2] = 0 \,, [X_1, X_3] = \mu X_3 \,, [X_3, X_2] = - \gamma X_2 \,.
\eEq
By appling the Jacobi identity we get $ \mu \gamma X_2 =0$ which, if we require $X_2\neq0$, is verified  either for  $\mu =0$ and $ \gamma \neq 0$ (see below \eqref{other dyon}) or for $\mu \neq0$ and $ \gamma =  0$ (see below \eqref{other dyon 2}), or for $\mu=0$ and $ \gamma= 0$ (trivial case of an abelian Lie algebra).

\bPr \label{new dyon} 
The case $X_4 = 0$ from the beginning, and with $X_8\neq 0$,  provides us with two base $3$-dyons.
\ePr

\bPf
\begin{enumerate}
\item the case with  $\mu =0$ and $ \gamma \neq 0$.

By putting   $X_8 = - \gamma X_2$,  and $X_4=X_5= X_6=X_7=0$ we get the $3$-dyon
\bEq\label{other dyon}
 [X_1, X_2] = 0 \,, [X_1, X_3] = 0 \,, [X_3, X_2] = - \gamma X_2 \,.
\eEq
The above dyon is incompatible with \eqref{dyon} while it is compatible with \eqref{3-dyon}.

\item the case $\mu \neq 0$ and $ \gamma =  0$.

We get the $3$-dyon
\bEq\label{other dyon 2}
 [X_1, X_2] = 0 \,, [X_1, X_3] = \mu X_3 \,, [X_3, X_2] = 0 \,.
\eEq

\end{enumerate}
\ePf

\bRm

It appears that the dyx-coaxial Lie algebra \eqref{dyx-family} can be assembled by one step from the case \eqref{3-dyon} and the latter one,  \eqref{other dyon 2}, by putting $\mu=\lam$,.
\eRm
We note in particular  that \eqref{incompatible} can not be seen as a dyx-family of  dyons since the two  dyons $[X_1, X_3] = \mu X_3$, $[X_3, X_2] = - \gamma X_2$ are incompatible and indeed if we apply the Jacobi identity we get particle-like structures made of single base dyons.

It seems therefore that the prolongation skeleton is homomorphic with quotient finite dimensional Lie algebras which have always the structure of a family of compatible dyons or single base $3$-dyons.
We note that the first dyon of \eqref{dyx-family} is compatible with \eqref{other dyon}, while the second dyon of \eqref{dyx-family} is not.

Summing up we were able to associate with the infinitesimal skeleton \eqref{open} a dyx-coaxial Lie algebra structure \eqref{dyx-family} and  particle-like Lie algebra structures made of three base $3$-dyons which are only partially compatible among them, \ie
\begin{itemize}

\item the first dyon of \eqref{dyx-family} is compatible with \eqref{other dyon}, while the second dyon of \eqref{dyx-family} is not. 
 
\item the first dyon of \eqref{dyx-family} is not compatible with \eqref{dyon}, while the second dyon of \eqref{dyx-family} is.

\end{itemize}

Note that \eqref{dyon}, \eqref{3-dyon}, \eqref{other dyon} and \eqref{other dyon 2} are not all compatible among them, even they are not compatible in triples, but they are only  compatible when took in couples.
\eRm

%---------------------------------------------------------------------------------------------------------------------------------------------------------------------------------
\subsubsection{A Lax pair associated with the dyx-coaxial Lie algebra structure of two compatible base  $3$-dyons}\label{Lax section}
%---------------------------------------------------------------------------------------------------------------------------------------------------------------------------------

Following a procedure similar to that of Section \ref{Conservation laws} we shall now derive a Lax pair  related to the Lie algebra \eqref{dyx-family}  to which the tower skeleton \eqref{commutator relations} is homomorphic. 

We refer again to equations \eqref{H} and \eqref{G}. The tower associated with case $2. (b)$ in Proposition \ref{dyonsProposition} becomes in this specific case
\bEq\label{H3}
H = e^{u}u_{z}X_1- u^2  e^{u} \ln u( \ln u -1 )[ X_1, [X_1,X_3 ] ] +  u  e^{u} \ln u [X_1,X_2] \,,
\eEq
\bEq\label{G3}
G = u_{x}X_1 - u(\ln u -1)[X_1,X_3]+X_2\,,
\eEq
By taking into account the representation $\rho$ given by relations \eqref{dyx-family}
we then get
\bEq\label{H3'}
H = e^{u}u_{z}X_1 -  \lam^2 u^2  e^{u} \ln u( \ln u -1 ) X_3  + \lam u  e^{u} \ln u  X_2 \,,
\eEq
\bEq\label{G3'}
G = u_{x}X_1 - \lam u(\ln u -1)  X_3 +X_2 \,,
\eEq

Now, let us represent the dyx-coaxial Lie algebra above in a space of `{\em pseudopotentials}' $\xi^k$ by
$X_1 = - \xi^1\der/\der\xi^1+\xi^2\der/\der\xi^2$, $X_2 = - \lam \xi^3\der/\der\xi^1$, $X_3 = \lam \xi^2\der/\der\xi^3$.

Again by sectioning the tower, equations \eqref{boh} and \eqref{boh2} provide the following
\beq\label{H2''}
& & e^{u} (u_{z}\xi^1 + \lam^2 u  \ln u \xi^3)   \der/\der\xi^1- e^{u}u_{z} \xi^2\der/\der\xi^2  + \\ & &+  \lam^3 u^2  e^{u} \ln u( \ln u -1 )  \xi^2\der/\der\xi^3    +  \xi^k_y \der/\der \xi^k= \xi^k_x \der/\der \xi^k  \,, \nonumber
\eeq
\beq\label{G2''}
& & - u_{x}( \xi^1 + \lam \xi^3 )\der/\der\xi^1+ u_{x}\xi^2\der/\der\xi^2  - \lam^2 u(\ln u -1)  \xi^2 \der/\der\xi^3 + \\ & &  + \xi^k_y\der/\der \xi^k =  \xi^k_z\der/\der \xi^k\,, \nonumber
\eeq
which gives us the inverse spectral problem
\bEq
\bxi_x= \bxi_y + {\hat \bM} \bxi \, \\
\bxi_z= \bxi_y + {\hat \bN} \bxi \,
\eEq
Where $\bxi =(\xi^1,\xi^2,\xi^3)^T$,  ${\hat \bM}$ and ${\hat \bN}$ are $3\times 3$ matrices such that $\hat{M}_{11}= e^{u} u_{z}$, $\hat{M}_{13}= e^u\lam^2 u  \ln u$, $\hat{M}_{22}= - e^{u}u_{z} $, $\hat{M}_{32} = \lam^3 u^2  e^{u} \ln u( \ln u -1 )$
$\hat{N}_{11} = -u_x$, $ \hat{N}_{13}=  - \lam u_x $, $ \hat{N}_{22} = u_x$, $ \hat{N}_{32} = - \lam^2 u(\ln u -1)$ and all the other entries are zeros. 
Here $\lam$ plays the role of a spectral parameter and, in view of \eqref{relations1}-\eqref{relations3} and \eqref{omega}, ${\hat \bM}$ and ${\hat \bN}$ can be considered a Lax pair related to the multidimensional Toda system  \eqref{1} (for spectral problems related to multidimensional nonlinear system see, \eg   \cite{Morris,Tondo}). This Lax pair should be compared with \cite{MaSa09}.

Compatibility of Lie algebraic structures being expressions of  compatibility of the corresponding Poisson structures, we note  here that Fernandes \cite{Fernandes93} studied the relationship between the master symmetries and bi-Hamiltonian structure of the Toda lattice. 
We stress that {\em dyons provide indeed particular examples of master symmetries} of related ordinary differential equations.

%-------------------------------------------------------
\subsection{Concluding remarks}
%-------------------------------------------------------

The structure of trix-coaxial and dyx-coaxial Lie algebras assembled in one step from couples of particle-like Lie algebra structures appears as an intrinsic feature of the Toda system \eqref{1}, at least associated with the chosen absolute parallelisms.
Indeed the similitude transformations seems to be the fundamental internal symmetries of the system (see \eg \cite{Alfinito1}).

As final remark, since \eqref{other dyon} is compatible with \eqref{3-dyon},  and since \eqref{dyon} is compatible with \eqref{other dyon 2}, we could construct the following dyx-coaxial Lie algebras:
\bEq\label{3-dyon 2}
[X_1, X_2]=\lam X_2\,, [X_1, X_3] = 0\,, [X_3, X_2]  = - \gamma X_2 \,,
\eEq
and 
\bEq\label{3-dyon 3}
[X_1, X_2]=0\,, [X_1, X_3] = \mu X_3\,, [X_2, X_3] = \nu X_3\,.
\eEq
However, it is important to realize that they could {\em not} be obtained from the skeleton \eqref{open} by the choice of an homomorphism, and therefore they are not identified as internal symmetries of the Toda system by the choice of the absolute parallelism given by Lemma \ref{lemma}.
The question if the choice of other forms of the absolute parallelism could identify them is open and will be the object of future investigations.

\section*{Acknowledgements}
Research partially supported by Department of Mathematics - University of Torino through the  projects PALM$\_$RILO$\_16\_ 01$ and FERM$\_$RILO$\_17\_ 01$ (MP) 
 and written under the auspices of GNSAGA-INdAM. 
The first author (MP) would like to acknowledge the contribution of the COST Action CA17139. The authors would like to thank M.V. Pavlov for bringing to their attention the reference \cite{Zakharov}.

%---------------------------------------------

\end{document}